\documentclass[aps,prb,twocolumn,superscriptaddress,floatfix,fleqn]{revtex4}

\usepackage{graphicx}
\usepackage{bm}
\usepackage{amsmath,amssymb}

\begin{document}


\title{All-optically transformable broad-band transparency and amplification in negative-index films
}

\author{Alexander  K. Popov}
\email{apopov@uwsp.edu}
\affiliation{Department of Physics \& Astronomy, University of
Wisconsin-Stevens Point, Stevens Point, WI 54481}

\author{Sergey  A. Myslivets}
\affiliation{Institute of Physics of the Russian Academy of
Sciences, 660036 Krasnoyarsk, Russian Federation}%

\date{\today}

\begin{abstract}
The possibility to produce laser-induced optical transparency of the metamaterial slab through the entire negative-index frequency domain is shown above the certain intensity threshold  of the control laser field.
\end{abstract}
\maketitle


\sloppy

Negative-index metamaterials (NIMs) exhibit highly unusual
optical properties and promise a great variety of unprecedented applications.\cite{Sh} However, strong optical absorption inherent to such materials imposes severe limitation on such applications. The possibility to overcome
this obstacle was proposed\cite{APB,OL} based on the coherent energy transfer from the control field to the signal through three-wave mixing, which is accompanied by optical parametric
amplification (OPA) in NIMs. It was shown that the transparency exhibits an extraordinary resonance behavior as a function of intensity of the control field and the NIM slab thickness. Basically, such resonances are narrow and the sample remains opaque anywhere  beyond the resonance field and sample parameters. Herein, we show that by adjusting the absorption indices for the signal and the idler, the slab can be made transparent and amplifying through the entire negative-index frequency domain and through a broad range of the slab thickness and intensity of the control fields provided that the the product of the intensity and the thickness remains above the certain threshold.

The basic idea of the proposal is as follows. We consider three coupled optical electromagnetic waves with wave vectors $k_j$ (j={1,2,3}) co-directed along the z axis.\cite{APB,OL}  The waves propagate through a slab of thickness L that possesses optical nonlinearity $\chi^{(2)}$. Here,  we assume magnetic nonlinearity. The outcomes do not change in the case of electric nonlinearity. Only two waves enter the slab, strong control field at $\omega_3$  and weak signal at $\omega_1$, which then generate a difference-frequency idler at $\omega_2=\omega_3-\omega_1$. The idler contributes back to the signal through the similar three-wave mixing process, $\omega_1=\omega_3-\omega_2$, and thus produce optical parametric amplification through the coherent energy transfer from the control field to the signal. The signal wave $H_1(\omega_1)$ is assumed backward due to negative refractive index  $n(\omega_1)< 0$. It means that the  energy flow $S_1$ is directed antiparallel to $k_1$ and, therefore, opposite to the z axis. Two other waves, the idler $H_2$  and the control field $H_3$, are ordinary waves with $k_{2,3}$ and $S_{2,3}$ directed along the  z axis since $n(\omega_2)> 0$ and $n(\omega_3)> 0$.  Consequently, the control wave enters the slab at z=0, whereas the signal at z=L. Generated idler travels along the z-axis. Unlike early proposals\cite{Har,Vol,Yar} and recent breakthrough experiment on backward-wave optical parametric oscillation in the sub-micrometer periodically poled nonlinear-optical (NLO) crystal\cite{Kh,Pas}, here all wave vectors are co-directed. This is crucially important because removes severe technical problem of phase matching  of the coupled waves with otherwise contra-directed wave vectors. The later is intrinsic to ordinary, positive index, materials.

The equations for the slowly-varying normalized amplitudes $a_{j}=\sqrt[4]{{\epsilon_{j}}/{\mu_{j}}}{h_{j}}/\sqrt{\omega_{j}}$ for the signal and idler waves can be written
in the form\cite{APB,OL}
\begin{eqnarray}
{da_{1}}/{dz} &=&-i{g}a_{2}^{\ast }\exp [i\Delta kz]+({\alpha_{1}}/{2}
)a_{1},  \label{a4} \\
{da_{2}}/{dz} &=&i{g}a_{1}^{\ast }\exp [i\Delta kz]-({\alpha_{2}}/{2 })a_{2}.
\label{a2}
\end{eqnarray}
Here, $g=(\sqrt{\omega_1\omega_2}/\sqrt[4]{\epsilon_1\epsilon_2/%
\mu_1\mu_2}) ({8\pi}/{c}){\chi^{(2)}}h_{3}$ is NLO coupling
coefficient, $h_{3}$ is amplitude of the control field, $\epsilon_j$ and $\mu_j$ are electric permittivity and magnetic permeability of the slab's material, phase mismatch $\Delta k=k_{3}-k_{2}-k_{1}$, $k_j$ are wave vectors directed along the z-axis, and $\alpha_{j}$
are the absorption  indices. The quantities $|a_{1,2}|^2$  are
proportional to the number of photons at the corresponding
frequencies. The
transmission factor for the negative-index signal, $T_1(z)=\left\vert {a_{1}(z)}/{a_{1L}}\right\vert ^{2}$,  and  energy conversion factor for the idler $\eta_2(z)=\left\vert {a_{2}(z)}/{a_{1L}}\right\vert ^{2}$, are derived from the solution to the coupled equations (\ref{a4}) and (\ref{a2}), whereas the control field is assumed homogeneous through the slab. Transmission is given by\cite{OL}:
\begin{equation}
T_1(z=0)=T_{10}=\left|\frac{\exp \left\{-\left[ \left( \alpha_{1}/2\right)-s
\right] L\right\}}{\cos RL+\left(s/R\right) \sin RL}\right|^2.  \label{T}
\end{equation}
Here, $s=({\alpha_{1}+\alpha_{2}})/({4})-i({\Delta k}/{2})$, and $R=\sqrt{
g^{2}-s^{2}}$. The fundamental difference
between the spatial distribution of the waves in ordinary and NI materials
is most explicitly seen in the limiting case $\alpha_j=0$, $\Delta k=0$. Then equation (\ref{T}) reduces to
\begin{equation}
T_{1}=1/[\cos(gL)]^2,  \label{T0}
\end{equation}
which depicts periodical oscillations of the transmission as the function of the  the slab thickness L and of the parameter g that is proportional to the control field strength. Transmission $T_{1}\rightarrow \infty$ at $g L \rightarrow (2j+1)\pi/2$, which indicates self oscillations.  Such  a  set of the geometrical transmission resonances (j=\{0, 1, 2,...\}), even at
$\Delta k =0$, is in a stark contrast with the  propagation properties in ordinary nonlinear-optical crystals. In the later case, for  all three wave vectors co-directed and $\Delta k =0$, the signal  and the idler would  exponentially grow across the slab  without any singularities as $|h_{1,2}|^2\propto exp(2gz)$. Such unusual transmission properties of the NIM slab occur  because the signal and the idler grow towards opposite directions, and they are determined by the boundary conditions on the opposite sides of the NIM slab.

\begin{figure}[!h]
\begin{center}
\includegraphics[width=.49\columnwidth]{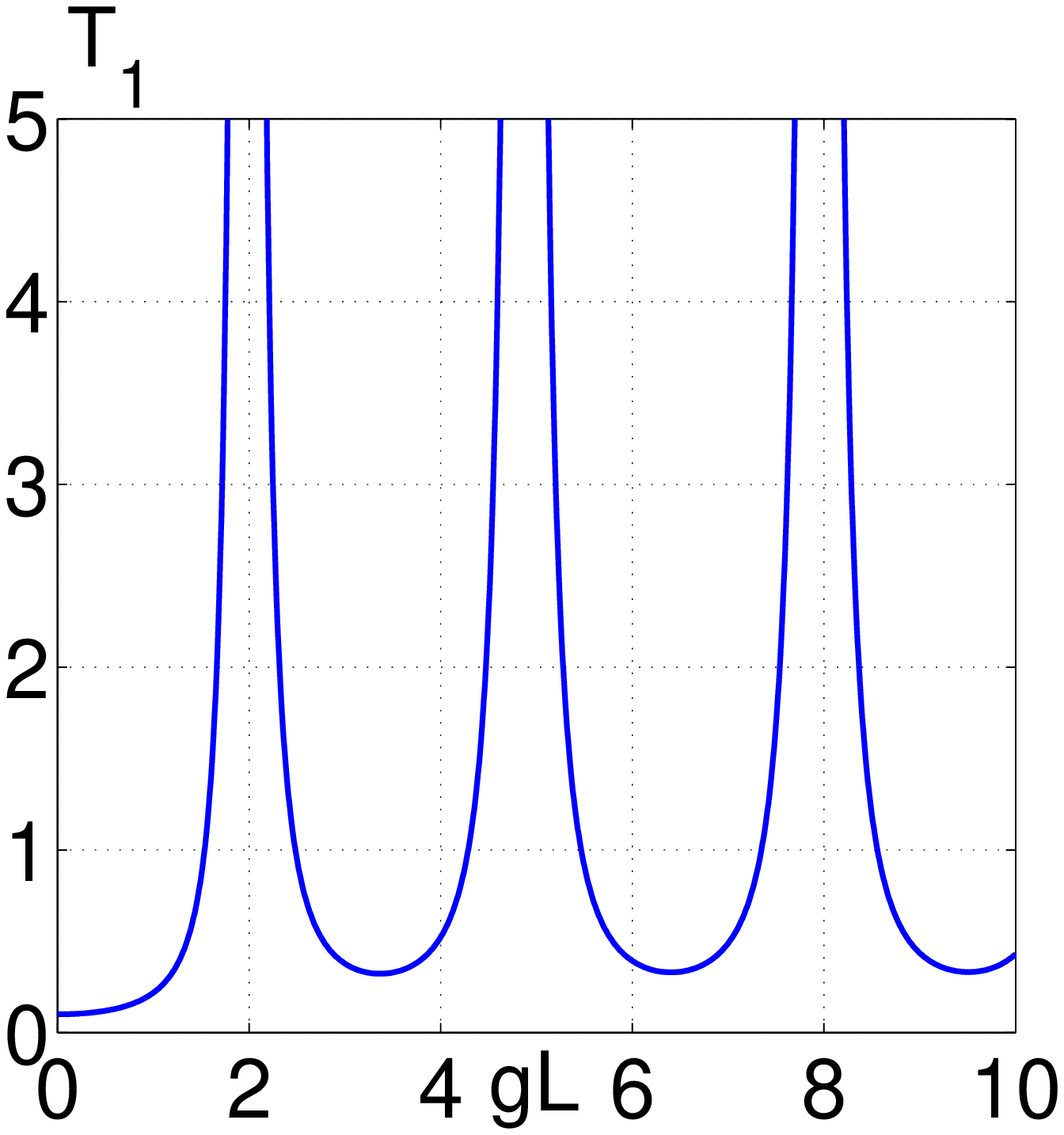}
\includegraphics[width=.49\columnwidth]{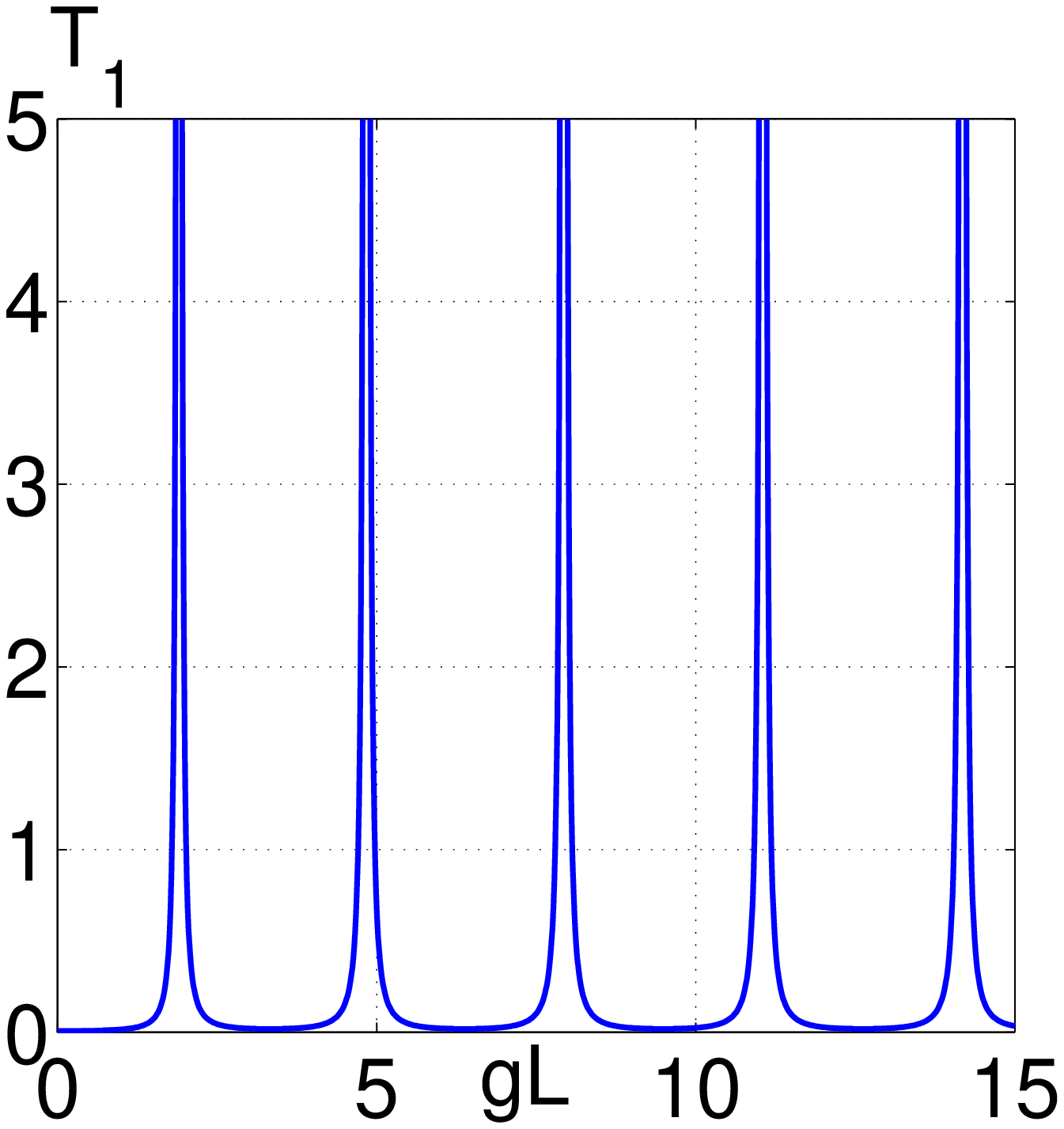}\\
(a)\hspace{35mm} (b)\\
\includegraphics[width=.49\columnwidth]{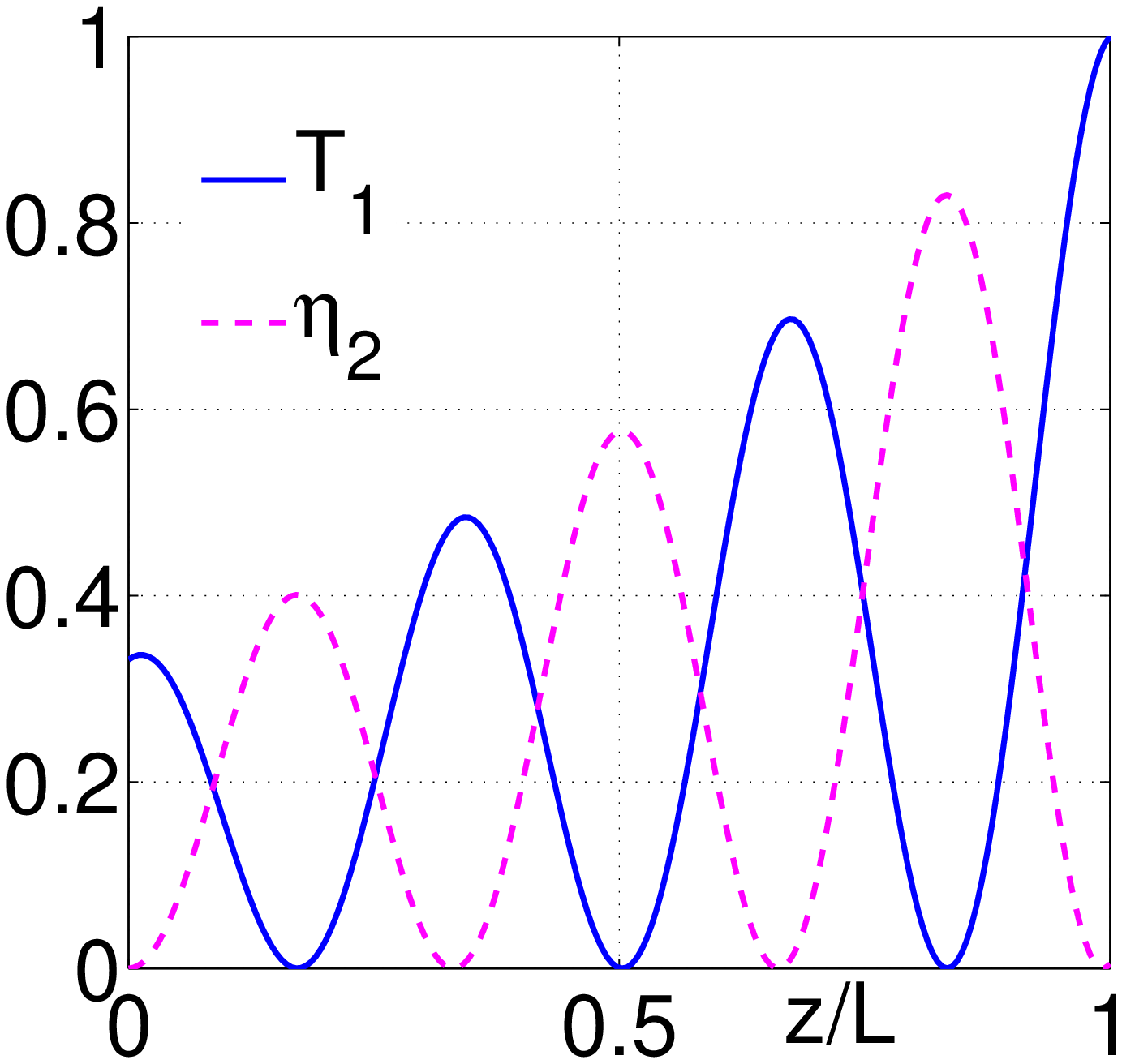}
\includegraphics[width=.49\columnwidth]{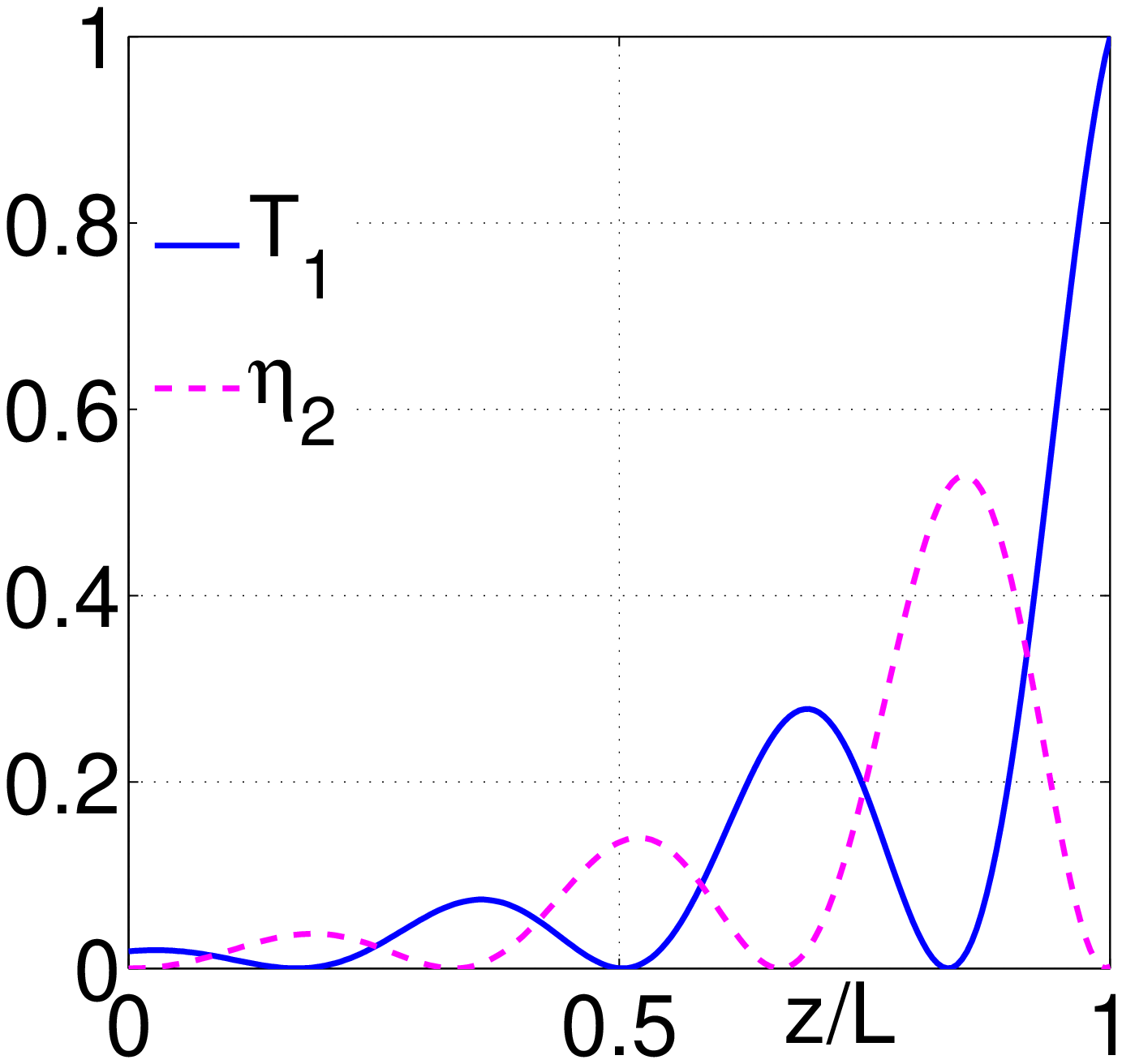}\\
(c)\hspace{35mm} (d)
\end{center}
\caption{\label{f1} Typical dependence of the output signal at z=0 on the intensity of the control field and the slab's thickness at different absorption indices for the signal and the idler, (a)-(b), and distribution of the fields inside the slab, (c)-(d). $\Delta k=0$.  (a) and (c): $\alpha_2L=0.1$, $\alpha_1L=2.3$. (b) and (d): $\alpha_2L=-3$, $\alpha_1L=5$. (c): gL=9.51; (d): gL=9.48.
}
\end{figure}
\begin{figure}[h!]
\begin{center}
\includegraphics[width=.49\columnwidth]{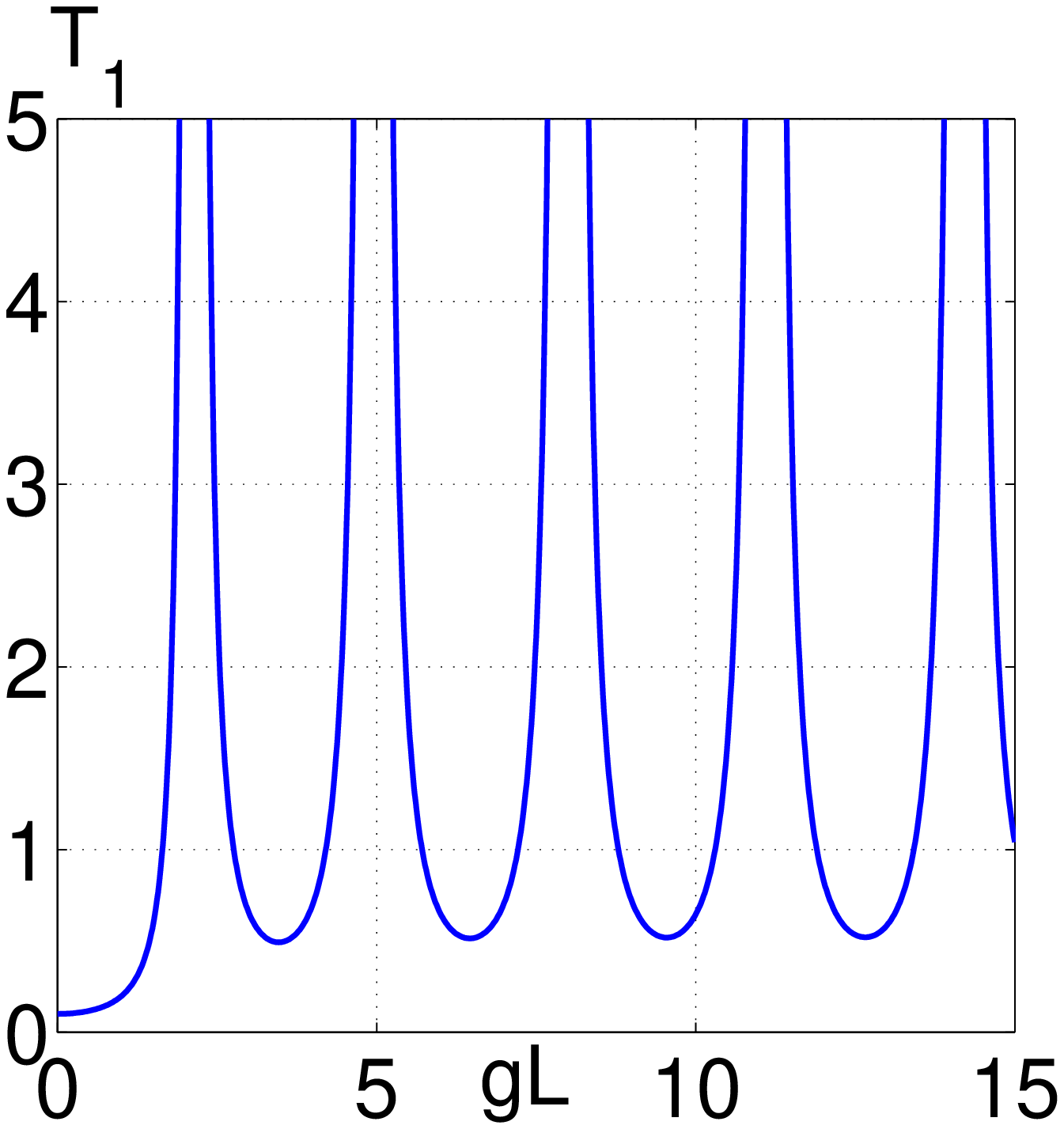}
\includegraphics[width=.49\columnwidth]{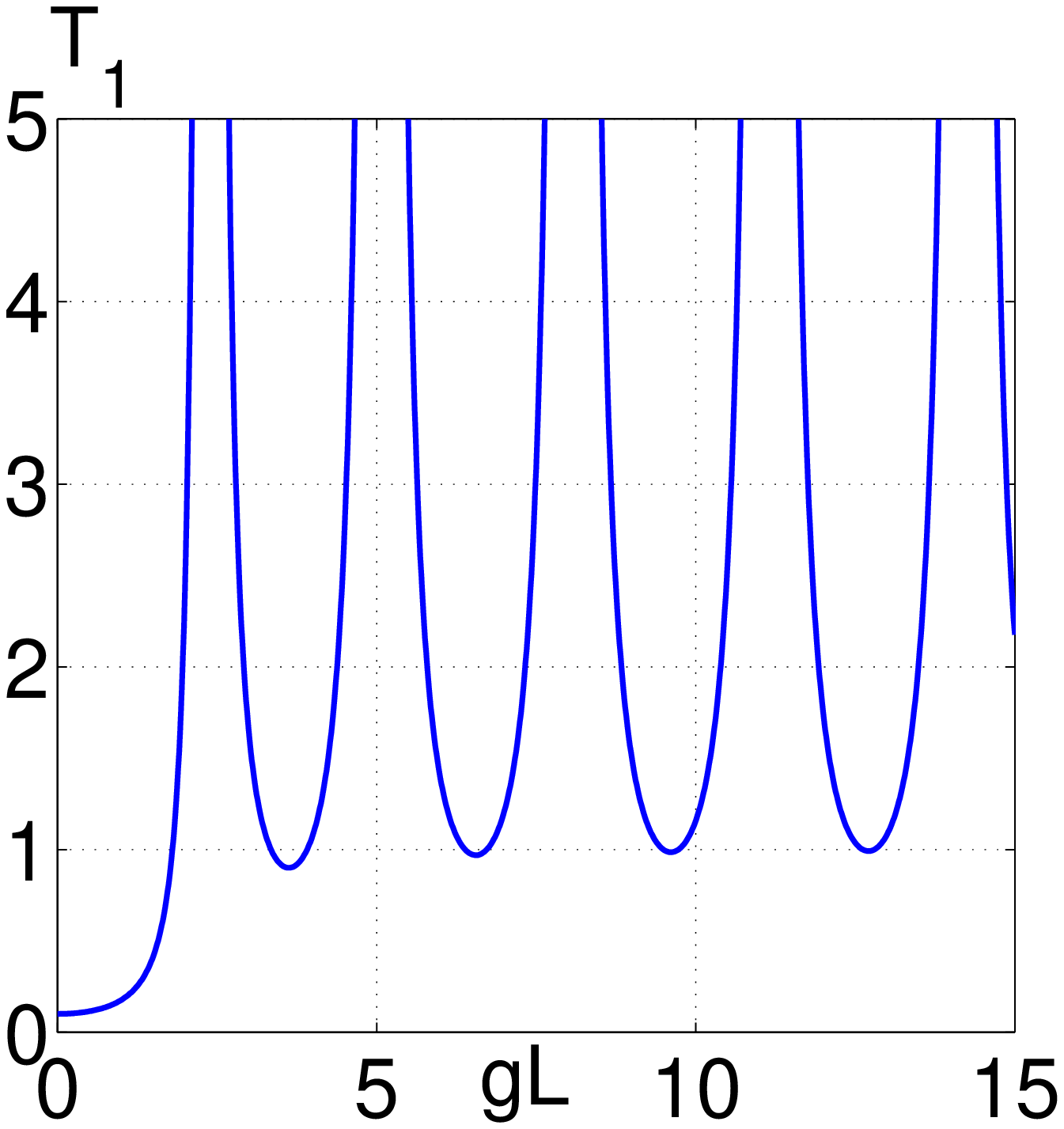}\\
(a)\hspace{30mm} (b)\\
\includegraphics[width=.49\columnwidth]{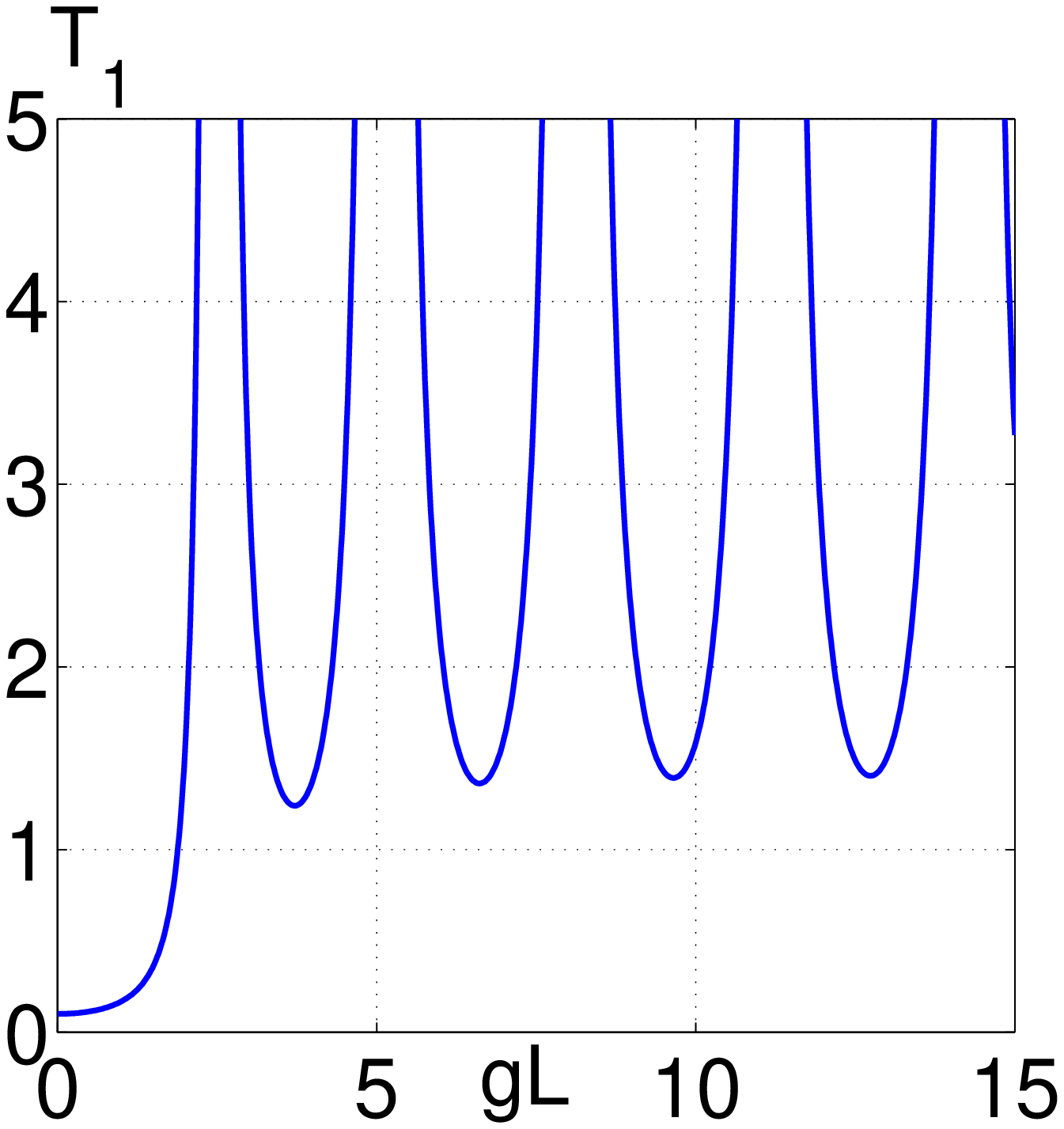}
\includegraphics[width=.49\columnwidth]{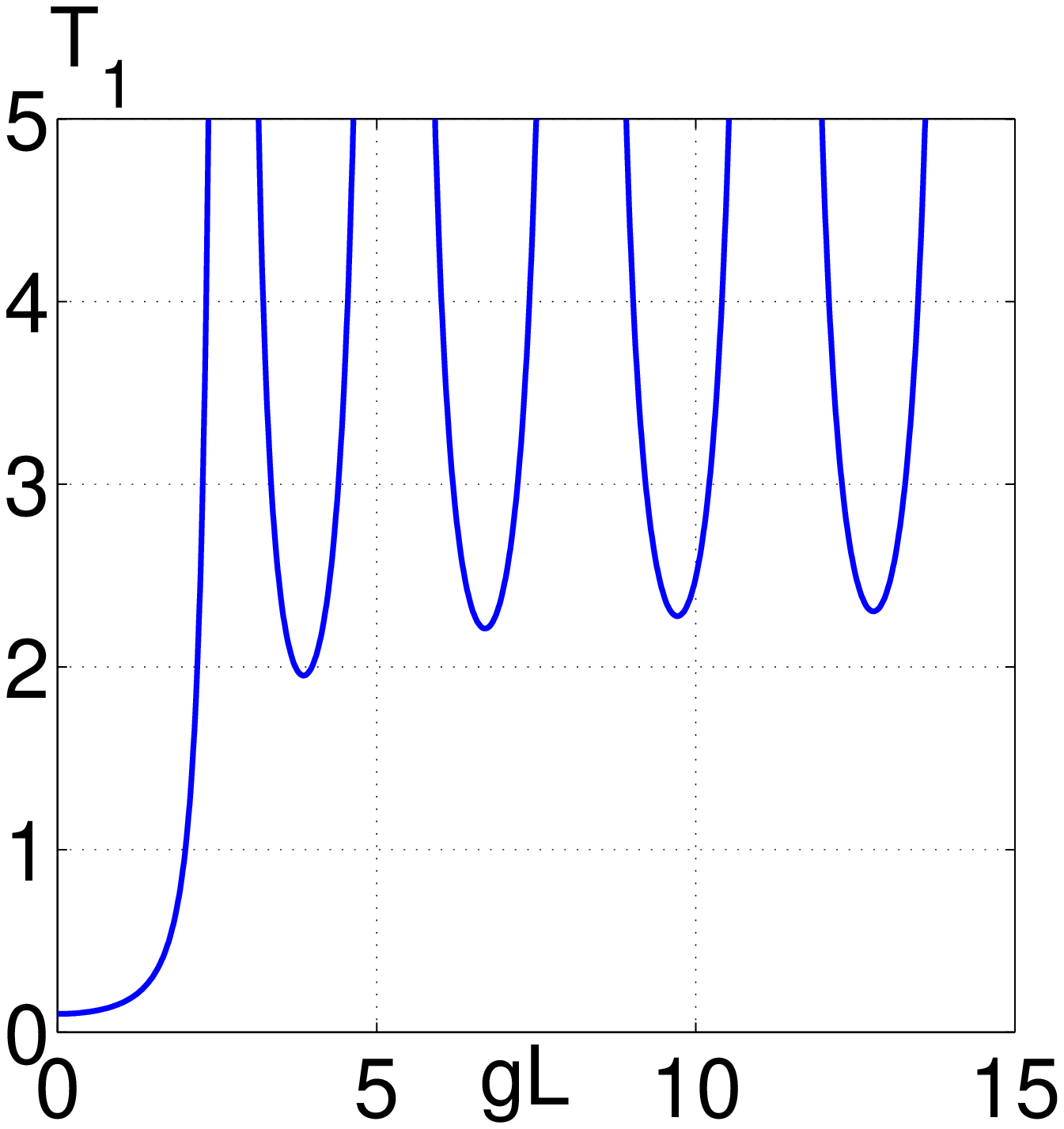}\\
(c)\hspace{30mm} (d)\\
\includegraphics[width=.49\columnwidth]{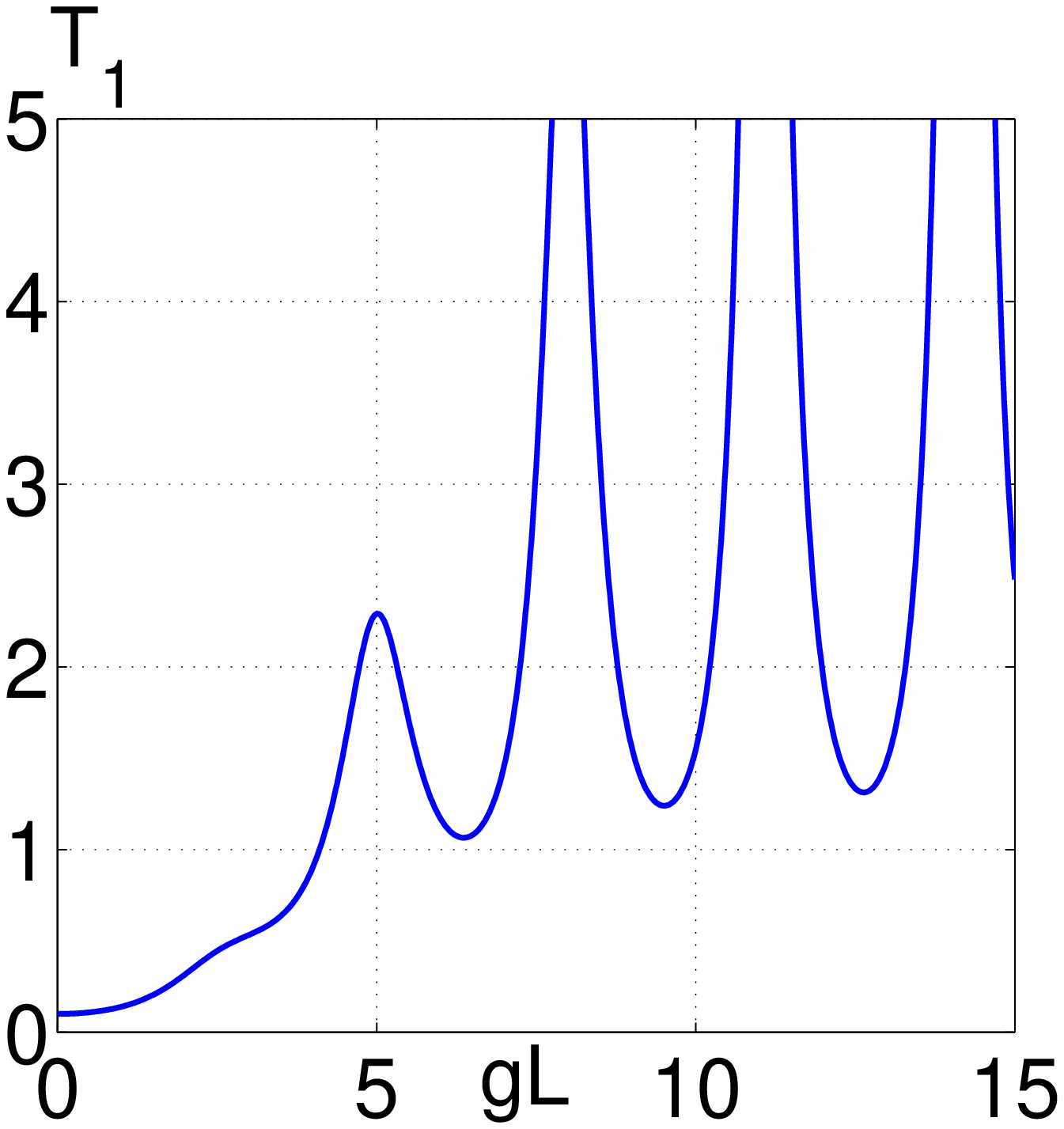}  \includegraphics[width=.49\columnwidth]{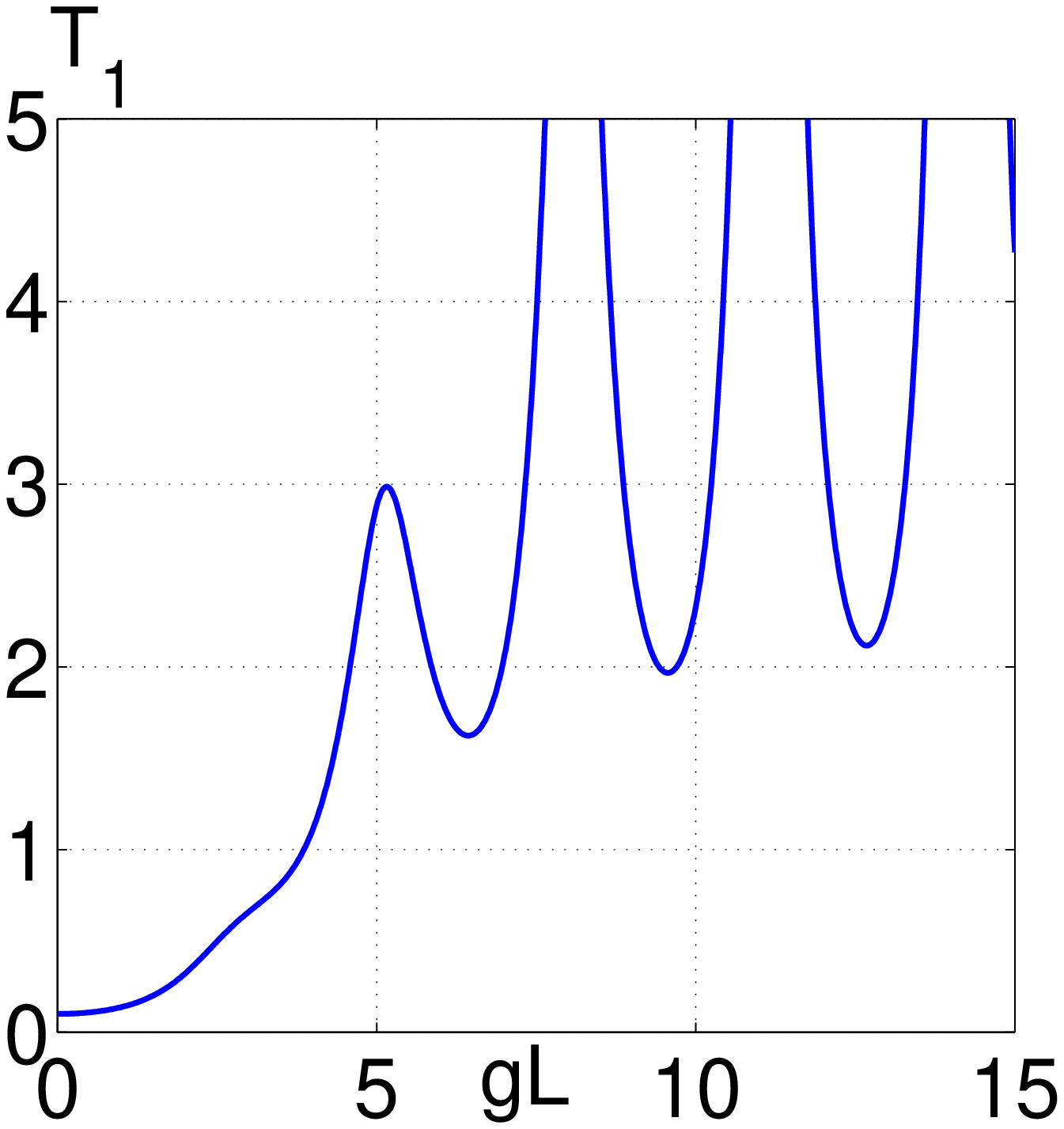}\\
(e)\hspace{30mm} (f)
\end{center}
\caption{\label{f2} Transmission of the NIM slab at $\alpha_1L=2.3$ and different values of $\alpha_2L$. (a)-(d): $\Delta kL=0$; (e)-(f): $\Delta kL=\pi$. (a): $\alpha_2L=1$; (b): $\alpha_2L=2.3$; (c) and (e): $\alpha_2L=3$; (d) and (f): $\alpha_2L=4$.
}
\end{figure}
Figures~\ref{f1} (a) and (b) presents numerical analysis of Eq. (\ref{T}). They depict a typical resonance dependence of the transmission factor for the originally strongly absorbing slabs on  parameter gL.  It is known that even weak amplification per unit length may lead to lasing provided that the corresponding frequency appears in a high-quality cavity or distributed feedback resonances. Such resonances are equivalent to a great extension of the effective length of a low-amplifying medium. Figures~\ref{f1} (a)-(b) demonstrate the similar behavior. The stronger is the signal absorption the more fine tuning to the OPA resonance is required. A \emph{strong resonance} dependence of the NIM slab transparency on the parameter gL indicates the necessity of \emph{fine tuning} for the intensity of the control field. This work is to show that such dependence can be transformed so that absolute transparency and amplification becomes robust and achievable through a \emph{wide range} of the parameter gL.
As seen from Eqs. (\ref{a4}) and (\ref{a2}), local NLO energy conversion rate for each of the waves is proportional to g-factor and to the amplitude of another coupled wave. Hence, the facts that the waves decay towards opposite directions have a significant influence on the entire NLO propagation process and on the entire transmission properties of the slab. As an example, Figs.~\ref{f1} (c)-(d) display unusual distributions of the fields inside the slab which correspond to the third transmission minimums. Oscillation amplitudes grow sharply with the approaching the resonances. Unless optimized, the signal maximum inside the slab may appear much greater than its output value at z=0. The comparison of Fig.~\ref{f1} (a) and (b)  suggests that the greater is difference between the signal and the idler absorption indices the more opaque is the slab beyond the resonances.
A typical NIM slab absorbs about 90\% of light at the frequencies which are in the NI frequency-range. Such absorption corresponds to $\alpha_1L\approx 2.3$. The slab becomes \emph{transparent within the broad range of the slab thickness and the control field intensity} if the transmission in the minimums is about or more than 1.
The examples of numerical analysis  given below are to prove the possibility to achieve such optical properties by the appropriate adjustment of the absorption indices. Figure~\ref{f2} depicts transmission properties of the NIM slab at $\alpha_1L=2.3$ and different magnitudes of the absorption index $\alpha_2L>0$ which are less,  equal and greater than $\alpha_1L$.
\begin{figure}[t!]
\begin{center}
\includegraphics[width=.49\columnwidth]{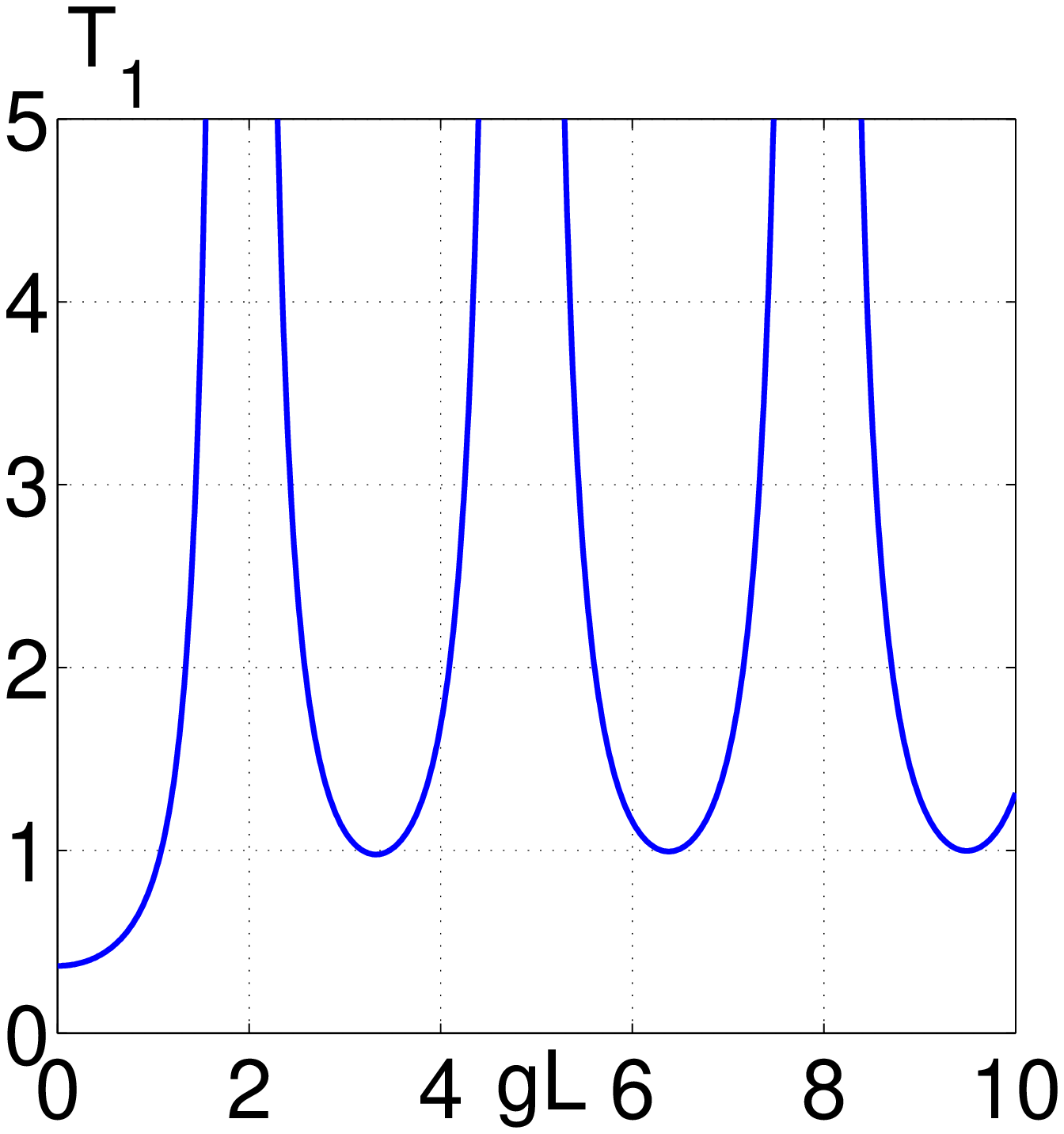}  \includegraphics[width=.49\columnwidth]{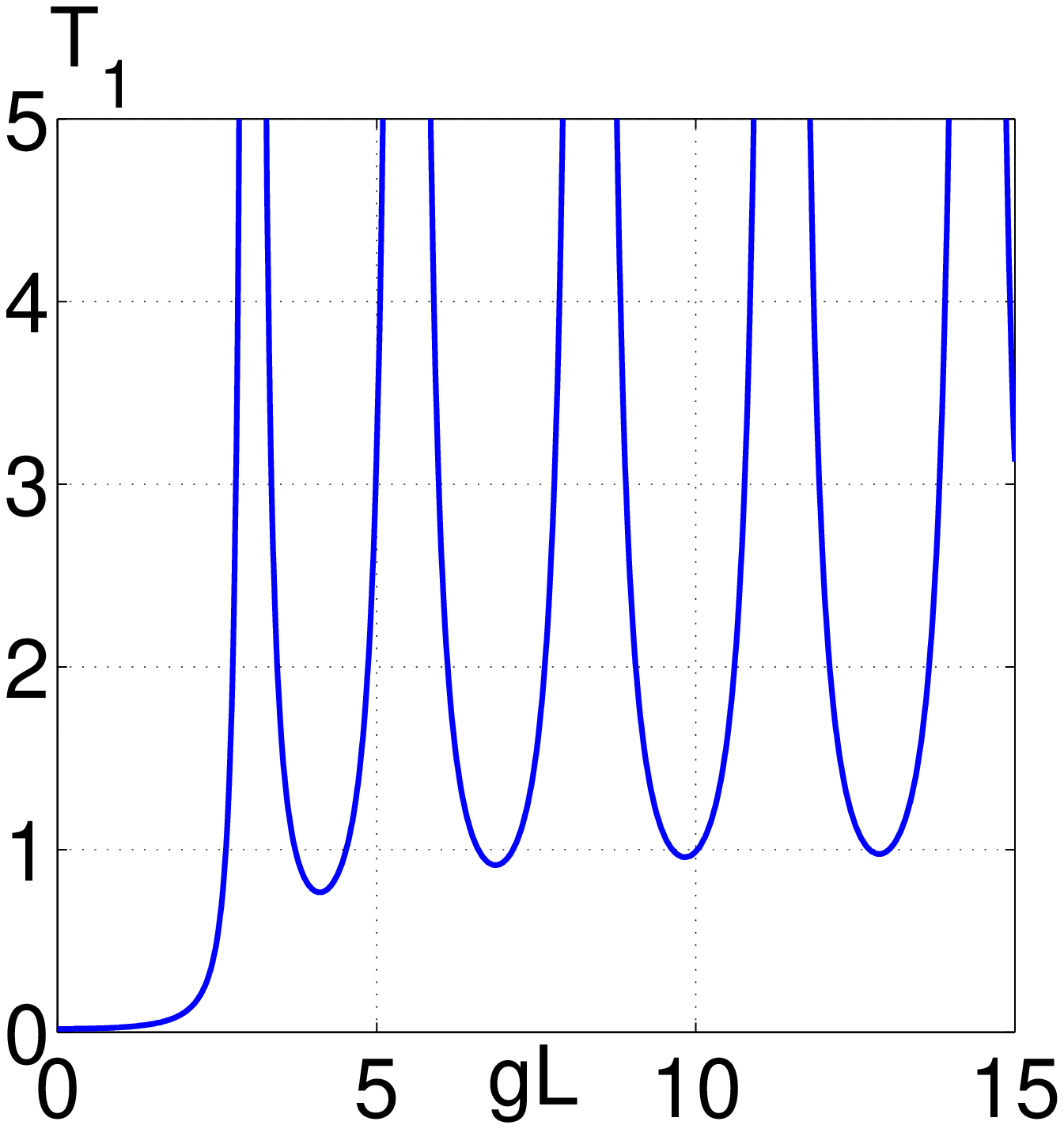}\\
(a)\hspace{30mm} (b)
\end{center}
\caption{\label{f3} Laser-induced transparency at different equal absorption indices for the signal and the idler. $\Delta k=0$. (a): $\alpha_2L=\alpha_1L=1$; (b): $\alpha_2L=\alpha_1L=4$.
}
\end{figure}
\begin{table}[t]
\caption{\label{T1} Dependence of the transmission minimums on the absorption index $\alpha_2L$. $\alpha_1L=2.3$}
\begin{tabular}{|c|c|c|c|c|c|c|}
\hline
& \multicolumn{3}{|c|}{$\alpha_2L=1$,  $\Delta k=0$}&\multicolumn{3}{|c|}{$\alpha_2L=2.3$, $\Delta k=0$}\\ \hline
$gL$ & 3.465 & 6.465 & 9.54 &3.63  &6.555  & 9.615 \\
\hline
$T_1$ &0.493 & 0.514 & 0.518 & 0.9 &0.969 & 0.986  \\
\hline
 & \multicolumn{3}{|c|}{$\alpha_2L=1$, $\Delta kL=\pi$}&\multicolumn{3}{|c|}{$\alpha_2=2.3$, $\Delta kL=\pi$}\\ \hline
  $gL$ &  & 6.24 & 9.405 &  &6.315 & 9.465 \\
  \hline
  $T_1$ & & 0.448& 0.487 &  &0.789  & 0.896  \\
  \hline
\end{tabular}
\begin{tabular}{|c|c|c|c|c|c|c|}
  \hline
   & \multicolumn{3}{|c|}{$\alpha_2L=3$,  $\Delta k=0$}&\multicolumn{3}{|c|}{$\alpha_2L=4$,  $\Delta k=0$}\\ \hline
  $gL$ & 3.72 & 6.6 & 9.66 &3.87  & 6.69 & 9.72 \\
  \hline
  $T_1$ &1.24 &1.362  &1.392  &1.953 &2.211 & 2.278  \\
  \hline
   & \multicolumn{3}{|c|}{$\alpha_2L=3$,  $\Delta kL=\pi$}&\multicolumn{3}{|c|}{$\alpha_2L=4$,  $\Delta kL=\pi$}\\  \hline
  $gL$ &  & 6.36 &9.51  &  &6.42 & 9.57 \\
  \hline
  $T_1$ & &1.065 &1.24  &  &1.624  &1.968   \\
  \hline
\end{tabular}
\end{table}
It is seen that the transmission in minimums becomes about or larger than 1 at  $\alpha_2\geq \alpha_1$.
Figure \ref{f3} presents transmission  at $\alpha_2=\alpha_1>2.3$. Being compared with Fig. \ref{f2} (b), it proves that
the transmission in the minimums depends rather on the ratio of the signal and the idler absorption indices then on their magnitude.
Transmission does not not drop below 1 both at low and high absorption indices provided that $\alpha_2\geq \alpha_1$ and gL-factor is larger than a  certain magnitude. The smaller are magnitudes of equal absorption indices, the closer to 1 is the transmission in the first minimum. It tends to 1 in the next minimums with the increase of the parameter gL at larger $\alpha_2= \alpha_1$. The resonances become narrower with the increase of the signal absorption index and experience a shift towards the larger magnitudes of the parameter gL.
Figures \ref{f2} (e) and (f) show that phase mismatch decreases transparency in the first maximums, but does not noticeably change transparency in the minimums so that it remains abut or larger than 1 if $\alpha_2\geq\alpha_1$. It is seen that the effect of phase mismatch decreases with the increase of the parameter gL.
Table \ref{T1} explicitly proves above presented conclusions regarding the changes in the positions and corresponding magnitudes of the transmission minimums.

To conclude, we have investigated the possibility to transform optical properties of the metamaterial slab in the negative-index frequency domain by the control laser. Strong absorption of negative-index electromagnetic waves is inherent to such materials.  It is crucially important for many applications of this revolutionary class of novel optical materials to ensure a robust transparency of a metamaterial slab for the negative-index signals. The basic idea of the proposed approach is coherent nonlinear-optical energy transfer from the control electromagnetic wave to the negative-index signal through a three-wave mixing and optical parametric amplification. We have reveal an extraordinary dependence of transmission of the negative-index backward-wave signal coupled with the ordinary, positive-index, control and idler waves on the ratio of their absorption indices. Such coupling scheme is intrinsic to negative-index metamaterials. With the aid of numerical simulations, we have shown the way to ensure nearly 100\%  transparency or amplification of the negative-index signal within a broad-range of intensities of the control fields and the slab's thicknesses. Among the counter-intuitive conclusions of the work is the recommendation to adjust the absorption index for the idler by increasing it above that for the signal. Thus, we have shown that the  minimum transparency can be transformed, increase and even turned into amplification so that it remains robust for any magnitudes of gL above a certain threshold.

We thank Vladimir  Shalaev for valuable discussions.  
This work was supported by the U.~S. Army Research Laboratory and by the U. S. Army Research Office under grant number W911NF-0710261.


\end{document}